\documentclass[12pt,a4paper]{article}
\usepackage{a4wide}
\usepackage{epsfig}
\usepackage{cite}
\usepackage{amsmath,amssymb}
\usepackage{graphicx}
\usepackage{latexsym}
\usepackage{array}
\usepackage{color}
\usepackage{float} 
\usepackage{wrapfig}
\usepackage{slashbox}
\newlength{\absize}
\newcommand{\half}{{\textstyle\frac{1}{2}}}

\def\lsim{\mathrel{\rlap{\raise 2.5pt \hbox{$<$}}\lower 2.5pt\hbox{$\sim$}}}
\def\gsim{\mathrel{\rlap{\raise 2.5pt \hbox{$>$}}\lower 2.5pt\hbox{$\sim$}}}

\newcommand{\Lumint}{{\cal L}_{\rm int}}

\newcommand{\hc}{\hbox {h.c.}}

\newcommand{\Meta}{M_{\eta^\pm}}
\newcommand{\GF}{G_\text{F}}

\allowdisplaybreaks 

\def\i11{{\mathbbm 1}}

\begin{document}

\thispagestyle{empty}
\renewcommand{\thefootnote}{\fnsymbol{footnote}}
\newpage\normalsize
\pagestyle{plain}
\setlength{\baselineskip}{4ex}\par
\setcounter{footnote}{0}
\renewcommand{\thefootnote}{\arabic{footnote}}
\newcommand{\preprint}[1]{%
\begin{flushright}
\setlength{\baselineskip}{3ex} #1
\end{flushright}}
\renewcommand{\title}[1]{%
\begin{center}
\LARGE #1
\end{center}\par}
\renewcommand{\author}[1]{%
\vspace{2ex}
{\Large
\begin{center}
  \setlength{\baselineskip}{3ex} #1 \par
\end{center}}}
\renewcommand{\thanks}[1]{\footnote{#1}}
\renewcommand{\abstract}[1]{%
\vspace{2ex}
\normalsize
\begin{center}
 \centerline{\bf Abstract}\par
 \vspace{2ex}
 \parbox{\absize}{#1\setlength{\baselineskip}{2.5ex}\par}
\end{center}}

\vspace*{4mm} 
\title{Phenomenology of charged scalars in the CP-Violating Inert-Doublet Model} 

\author{
P. Osland\thanks{Per.Osland@ift.uib.no} \\
Department of Physics and Technology, University of Bergen, \\
Postboks 7803, N-5020  Bergen, Norway}
\author{A. Pukhov\thanks{Pukhov@lapp.in2p3.fr} \\
Skobeltsyn Inst.\ of Nuclear Physics, Moscow State Univ., Moscow 119991, Russia}
\author{G. M. Pruna\thanks{Giovanni\_Marco.Pruna@tu-dresden.de} \\
TU Dresden, Institut f\"ur Kern- und Teilchenphysik,
Zellescher Weg 19, D-01069 Dresden, Germany}
\author{M. Purmohammadi\thanks{mpu087@ift.uib.no} \\
Department of Physics and Technology, University of Bergen, \\
Postboks 7803, N-5020  Bergen, Norway}

%
%
\abstract{
We study the production and decay of charged scalars, $\eta^\pm$, in the context of a CP-Violating Inert-Doublet Model. 
The model is an extended version of the Inert Doublet Model with an extra Higgs doublet and provides new sources of CP violation and a dark matter candidate. As compared with the 2HDM,
the  particle spectrum contains two additional neutral scalars and a charged pair. These particles are subject to a $Z_2$ symmetry, but can be pair-produced in hadronic collisions. If a charged scalar is included in the pair, it decays to the stable dark-matter candidate (i.e., the lightest neutral inert scalar) plus Standard Model matter that consists of either two jets or a single lepton (from a virtual or real $W$ or $Z$) plus missing transverse energy. Since the single production channel is available only at hadronic colliders, we consider the Large Hadron Collider environment, hence we discuss experimental perspectives and possible hallmarks of the model, such as events with a displaced vertex.
}


\setcounter{footnote}{0}
\renewcommand{\thefootnote}{\arabic{footnote}}

\section{Introduction}
The Standard Model (SM) of particle physics is an effective theory that describes the present collider experiments to a remarkable accuracy, but there are empirical reasons which suggest that it can not be an ultimate theory of nature. Cosmological evidence shows that about $23\%$ of the energy density of the Universe is in the form of dark matter \cite{Bertone:2004pz} and we still have no idea what it consists of. Indeed, the 
identification of the dark matter particle and its production mechanism are among the most challenging problems of astroparticle physics today. Over the last few decades the paradigm of dark matter candidates has shifted towards particle dark matter. Particle candidates are proposed in several extensions for physics beyond the Standard Model and new spin-zero scalars are among those that have received particular attention.

The Inert Doublet Model (IDM) is a minimal extension of the Standard Model which could account for the Dark Matter (DM). The SM particle content is augmented by an extra weak scalar SU(2) doublet, which is odd under an unbroken $Z_2$ symmetry, rendering the lightest member stable. This is assumed to be a neutral scalar, denoted $S$, and is the dark matter candidate. The model has been introduced and  studied in different contexts\cite{Deshpande:1977rw,Cirelli:2005uq,Barbieri:2006dq,LopezHonorez:2006gr,Hambye:2007vf,Cao:2007rm, Andreas:2008xy, Lundstrom:2008ai,Hambye:2009pw,LopezHonorez:2010tb}.

The scalar spectrum of the model has 
another neutral particle, $A$, and a pair of charged ones.
These particles can all be produced at colliders via their couplings to 
electroweak gauge bosons (and the Higgs boson), subject to the $Z_2$ symmetry. It was first introduced to provide a mass mechanism for neutrinos and it could also alleviate the Little hierarchy by allowing higher values of mass for Higgs particle, though the latter argument is now irrelevant in view of the recent discovery of a Higgs signal around 125~GeV \cite{atlas:2012gk,cms:2012gu}.
 
It is evident that the introduction of CP violation in
the scalar sector would make the model more interesting, therefore an
extension to a Two-Higgs-Doublet Model (2HDM) plus an inert doublet
model was proposed \cite{Grzadkowski:2009bt} and its viable parameter 
space was explored  \cite{Grzadkowski:2010au}.  We shall refer to the
resulting structure as the IDM2. The IDM2 allows two interesting mass regions for the dark matter candidate ($m\sim m_W$ and $m\gsim 500~\text{GeV}$). The low-mass region, with dark matter in the range 10 to 60~GeV, is ruled out by the XENON100 results \cite{Aprile:2011hi,Aprile:2012nq,Grzadkowski:2011zn}.

In this paper our aim is to study the production mechanisms of the associated charged scalars and their decay at the LHC. At the same time we will confront the model with the latest experimental limits, in particular, those from the LHC \cite{atlas:2012gk,cms:2012gu} and XENON100 \cite{Aprile:2011hi,Aprile:2012nq}. We will see that if the scalar particles are produced at the LHC  they could leave some signals which may enable one to discover them.

Then, we examine the main production mechanisms involving inert charged scalars (single and double production) and the favoured decay modes. Among all of them, we select the single production as it is a channel that can only be produced at hadronic colliders. Hence, we discuss in detail the production of a charged scalar together with a DM particle, with the former decaying to two jets via a virtual $W$. Because of detector effects, the di-jet final state will be collected at the LHC as an ``effective'' single jet, implying the study of the process $pp\to\text{ jet}+\text{ MET}$.

In view of the facts that the low-mass region is basically ruled out, and that the high-mass region ($M_{\eta^\pm}>M_S\gsim550~\text{GeV}$ is experimentally hard to explore (low cross sections), we shall focus on the case of $M_S=75~\text{GeV}$.

The paper is organized as follows. In the next section we briefly present the model and its particle content. Then, in section~\ref{Sec:parameters} we introduce constraints that will be imposed, in section~\ref{Sec:eta-properties} we discuss the charged-scalar properties, in section~\ref{Sec:allowed-parameters} we outline the allowed parameter regions. Then, in section~\ref{Sec:LHC} we   determine the cross section for a set of viable parameters, and in section~\ref{Sec:experimental} we discuss experimental possibilities. Finally, section~\ref{Sec:summary} contains a brief summary.

\section{The IDM2}
\label{Sec:def-model}
\setcounter{equation}{0}
\subsection{Fields and potential}
The IDM2 may be seen as a type-II version of the 2HDM augmented by an inert SU(2) doublet that provides a dark matter candidate.  We denote the doublets of the 2HDM as
\begin{equation}
\Phi_1=\left(
\begin{array}{c}\varphi_1^+\\ (v_1+\eta_1+i\chi_1)/\sqrt{2}
\end{array}\right), \quad
\Phi_2=\left(
\begin{array}{c}
\varphi_2^+\\ (v_2+\eta_2+i\chi_2)/\sqrt{2}
\end{array}
\right),
\label{Eq:basis}
\end{equation}
where $v^2=v_1^2+v_2^2$ and $\tan\beta=v_2/v_1$.
The inert doublet is decomposed as
\begin{equation}
\eta = \left(
\begin{array}{c}
 \eta^+ \\ (S+iA)/\sqrt{2}
\end{array}
\right),
\label{eta}
\end{equation}
it transforms under an unbroken $Z_2$ symmetry as $\eta
\to -\eta$ which ensures that $\eta$ couples only bilinearly to other
scalars and to the gauge sector. All other fields remain neutral under
this transformation.

The scalar couplings will be given by
\begin{equation} \label{Eq:fullpot}
V(\Phi_1,\Phi_2,\eta)
= V_{12}(\Phi_1,\Phi_2) + V_3(\eta) + V_{123}(\Phi_1,\Phi_2,\eta)
\end{equation}
where the 2HDM and inert-sector potentials read
\begin{align}
V_{12}(\Phi_1,\Phi_2) &= -\frac12\left\{m_{11}^2\Phi_1^\dagger\Phi_1
+ m_{22}^2\Phi_2^\dagger\Phi_2 + \left[m_{12}^2 \Phi_1^\dagger \Phi_2
+ \hc\right]\right\} \nonumber \\
& + \frac{\lambda_1}{2}(\Phi_1^\dagger\Phi_1)^2
+ \frac{\lambda_2}{2}(\Phi_2^\dagger\Phi_2)^2
+ \lambda_3(\Phi_1^\dagger\Phi_1)(\Phi_2^\dagger\Phi_2) \nonumber \\
&+ \lambda_4(\Phi_1^\dagger\Phi_2)(\Phi_2^\dagger\Phi_1)
+ \frac12\left[\lambda_5(\Phi_1^\dagger\Phi_2)^2 + \hc\right],
\label{v12} \\
V_3(\eta) &= m_\eta^2\eta^\dagger \eta + \frac{\lambda_\eta}{2}
(\eta^\dagger \eta)^2.
\label{v3}
\end{align}
In order to keep the number of parameters at a manageable level, we
impose ``dark democracy'', the inert doublet has the same interaction with $\Phi_1$ as with $\Phi_2$,
\begin{align}
V_{123}(\Phi_1,\Phi_2,\eta)
&=
\lambda_{a} (\Phi_1^\dagger\Phi_1)(\eta^\dagger \eta)
+\lambda_{a} (\Phi_2^\dagger\Phi_2)(\eta^\dagger \eta) \nonumber  \\
& +\lambda_{b}(\Phi_1^\dagger\eta)(\eta^\dagger\Phi_1)
+\lambda_{b}(\Phi_2^\dagger\eta)(\eta^\dagger\Phi_2) \nonumber  \\
&
+\half\left[\lambda_{c}(\Phi_1^\dagger\eta)^2 +\hc \right]
+\half\left[\lambda_{c}(\Phi_2^\dagger\eta)^2 +\hc \right],
\label{v123}
\end{align}
Furthermore, we take $\lambda_c$ to be real.

The introduction of additional scalars is accompanied by the possibility of having unacceptable flavour-changing neutral currents (FCNC) and to remove them at the tree level it is assumed that the total Lagrangian is symmetric under a $Z'_2$  symmetry which transforms $\Phi_1\to -\Phi_1$ and $u_R\to -u_R$  and leaves all other fields unchanged. This symmetry is however broken by the $m_{12}^2$ term.

In Eq. (\ref{v12}), $\lambda_5$ and $m_{12}^2$ could be complex, allowing for CP violation. 
As a consequence of the unbroken $Z_2$ associated with the inert doublet, and with respect to the assumption that the inert doublet does not develop a vacuum expectation value, $\langle\eta\rangle=0$, there is no mixing in the mass matrix between $\Phi_{1,2}$ and $\eta$\cite{Grzadkowski:2009bt}. Since  $\eta^\pm$ decouples from $G^\pm$ and $H^\pm$, there is no
CP-violation mediated by charged scalars. However CP is violated in the neutral non-inert scalar sector in the same way as in the 2HDM. For recent reviews of the allowed parameter space of the 2HDM sector, see Refs.~\cite{WahabElKaffas:2007xd,Basso:2012st}.

The dark-sector masses can be written as:
\begin{align}
M^2_{\eta^\pm}
&=m_\eta^2+\half\lambda_a\,v^2, \nonumber \\
M^2_S
&=m_\eta^2+\half(\lambda_a+\lambda_b+\lambda_c)v^2
=M^2_{\eta^\pm}+\half(\lambda_b+\lambda_c)v^2, \nonumber \\
M^2_A
&=m_\eta^2+\half(\lambda_a+\lambda_b-\lambda_c)v^2
=M^2_{\eta^\pm}+\half(\lambda_b-\lambda_c)v^2,
\label{inmass}
\end{align}
where $m_\eta$ is a mass parameter of the inert 
potential. We shall take the scalar, $S$, to be the DM
particle, i.e., $M_S<M_A$.  The other choice would simply correspond
to $\lambda_c\to-\lambda_c$, without any modification of the phenomenology described here.

The relations (\ref{inmass}) can be reformulated as
\begin{subequations} 
\label{Eq:lambda-vs-splitting}
\begin{align}
\lambda_a&=\frac{2}{v^2}
\left(M^2_{\eta^\pm}-m_\eta^2\right), \\
\lambda_b&=\frac{1}{v^2}\left(M^2_S+M^2_A-2M^2_{\eta^\pm}\right), \\
\lambda_c&=\frac{1}{v^2}\left(M^2_S-M^2_A\right).
\end{align}
\end{subequations}
Thus, these couplings of the inert doublet to the non-inert Higgs sector can be expressed in terms of the mass splittings and the soft mass parameter $m_\eta$.

The coupling of the inert particles to the Higgs sector is largely controlled by \cite{Grzadkowski:2009bt,Grzadkowski:2010au}
\begin{equation} \label{Eq:lambda_L}
\lambda_L\equiv \half(\lambda_a+\lambda_b+\lambda_c)
=\frac{M_S^2-m_\eta^2}{v^2}.
\end{equation}
Thus, the splitting $M_S^2-m_\eta^2$ (which may be positive or negative) is a measure of this coupling strength.

\subsection{Particle content}
The particle content of the model can be organized into two sectors:
\begin{itemize}
\item
The familiar Higgs scalars of the 2HDM, being three neutral ones, $H_1$, $H_2$, $H_3$ ($M_1<M_2<M_3$), and a charged pair, $H^\pm$ ($M_{H^\pm}\gsim380~\text{GeV}$). In a CP-conserving limit (there are three such limits) one of the neutral ones would be CP-odd, usually demoted $A$. We shall here not consider such limits.
\item
The inert sector contributes two neutral ones, denoted $S$ (DM candidate) and $A$, and a charged pair, $\eta^\pm$ ($M_{\eta^\pm}\gsim70~\text{GeV}$). (This $A$ is different from the one mentioned abovee, as a limiting case of one of the $H_i$.)
\end{itemize}
We assume type-II Yukawa couplings, hence the mass of the charged Higgs particle can not be too low, $M_{H^\pm}\gsim380~\text{GeV}$, due to the $b\to s\gamma$ constraint \cite{Hermann:2012fc}.

Most of the time, when we refer to ``charged scalars'', we will refer to those of the inert sector, since they could be significantly lighter (and possibly more easily produced) than the charged Higgs, $H^\pm$. The reason they can be lighter, is that they have no Yukawa couplings, and are not affected by the $b\to s\gamma$ constraint.
\section{The parameter space}
\label{Sec:parameters}
\setcounter{equation}{0}
The parameter space of the potential is subject to a variety of theoretical and experimental constraints.  We have considered two sets of relevant constraints in order to find allowed regions.
\subsection{Theoretical constraints}
We impose:
\begin{itemize}
\item
Positivity---the potential must be positive for asymptotic values of the fields involved.
\item
Perturbative unitarity---Higgs-Higgs scattering amplitudes are constrained.
\item
Perturbativity---the individual $\lambda$'s of the potential are constrained.
\item
Global minimum---the adopted minimum of the potential is the deepest one.
\end{itemize}
These constraints are imposed in the same manner as in Refs.~\cite{Grzadkowski:2009bt,Grzadkowski:2010au}, where relevant references can be found. The last constraint is computationally expensive, and thus imposed last of all.

\subsection{Observational constraints}
\label{Subsect:obs-constraints}

Another set of constraints  stems from experimental observations. The following are taken into account:
\begin{itemize}
\item {\bf  General constraints from the charged Higgs boson.} These mostly arise from B physics and are only weakly dependent on the neutral-Higgs sector. The most relevant ones are the $b\rightarrow s\gamma$ transition, the $B\rightarrow \tau \bar{\nu_\tau} X$ branching ratio and $B-\bar{B}$ oscillations. For a treatment of these, see \cite{Grzadkowski:2009bt,Grzadkowski:2010au}.

\item {\bf Constraints on the neutral scalars.} These are dependent on the neutral-Higgs sector. Important ones are the measurement of  the branching ratio of $Z\rightarrow b\bar{b}$ ($R_b$), and (since we allow CP violation) the electron electric dipole moment.
There are also bounds from the electroweak precision data on the oblique parameters $T$ and $S$
\cite{Peskin:1990zt,Grzadkowski:2009bt,Grzadkowski:2010au}.

\item {\bf Neutral Higgs searches at the LHC.}
We require the parameter space to meet two conditions: 
\begin{itemize}
\item The production and subsequent decay of a neutral Higgs to $\gamma\gamma$, around $M=125~\text{GeV}$ is taken to be within a factor of 2 from the Standard Model. Assuming the dominant production to be via gluon fusion, this can be approximated as $0.5\leq R_{\gamma\gamma}\leq2$, where we define
\begin{equation}
R_{\gamma\gamma}=\frac{\Gamma(H_1\to gg)\text{BR}(H_1\to\gamma\gamma)}
{\Gamma(H_\text{SM}\to gg)\text{BR}(H_\text{SM}\to\gamma\gamma)}.
\end{equation}
As in \cite{Basso:2012st}, we take into account the modified couplings of $H_1$ to the $t$-quark (from both the scalar and pseudoscalar components of $H_1$) in the loop on the production side, and to the modified $W$ and fermion contributions on the $\gamma\gamma$ side, as well as the $H^\pm$ contribution. In addition, there is a contribution from an $\eta^\pm$ loop.

\item The production and subsequent decay, dominantly via $ZZ$ and $WW$, is constrained in the mass range $130~\text{GeV}\lsim M \lsim600~\text{GeV}$. We consider the quantity
\begin{equation}
R_{ZZ}=\frac{\Gamma(H_j\to gg)\text{BR}(H_j\to ZZ)}
{\Gamma(H_\text{SM}\to gg)\text{BR}(H_\text{SM}\to ZZ)},
\end{equation}
for $j=2,3$ and require it to be below the stronger 95\% CL obtained by ATLAS or CMS. This constraint thus affects the product of Yukawa and gauge couplings of $H_2$ and $H_3$. For the total widths of $H_2$ and $H_3$ we also include $H_j\to H_1H_1$ and $H_j\to ZH_1$.
\end{itemize}

\item {\bf  Inert-sector constraints.}
We have adopted the following bound on the dark matter relic density \cite{Hinshaw:2008kr}
\begin{equation}
\Omega_\text{DM} h^2=0.1131\pm0.0034.
\end{equation}
We estimate the amount of dark matter from an
implementation of {\tt micrOMEGAs}
\cite{Belanger:2006is,Belanger:2008sj}.
For the heavier, neutral member of the inert sector ($A$), we adopt the bound obtained from a re-analysis of LEP data \cite{Lundstrom:2008ai}, approximated as $M_A>110~\text{GeV}$. For the charged member, we adopt the LEP bound on the chargino mass \cite{Pierce:2007ut}, $M_{\eta^\pm}>70~\text{GeV}$, which is slightly more conservative than the bound on charged Higgs bosons, $M_{H^\pm} > 79.3~\text{GeV}$, adopted by Ref.~\cite{LopezHonorez:2006gr}.
As will be discussed below, we take $M_S$ in the region compatible with the XENON100 results \cite{Aprile:2011hi,Aprile:2012nq}.
\end{itemize}

\section{Charged scalar properties}
\label{Sec:eta-properties}
\setcounter{equation}{0}

Here, we review the charged-scalar properties. For completeness, we remark that those of the unstable neutral one would be analogous \cite{Grzadkowski:2009bt,Grzadkowski:2010au}.
\subsection{Charged scalar production}

At the LHC, due to the $Z_2$ symmetry imposed on the potential, the charged scalars of the inert sector could be (a) directly pair produced, (b) produced in association with a neutral one, or (c,d) as a decay product from a heavier, neutral, one:
\begin{subequations}\label{eq:prod}
\begin{align}
pp&\to \eta^+\eta^- X, \\
pp&\to S\eta^\pm X, A\eta^\pm X, \\
pp&\to SAX\to S\eta^\pm X', \\
pp&\to AAX\to SS\eta^\pm\eta^\pm X'.
\end{align}
\end{subequations}
The relative importance of these processes will of course depend on the spectrum. For example, the last two will only be relevant if $M_{\eta^\pm}<M_A$. 

\begin{figure}[ht]
\begin{center}
\includegraphics[scale=0.9]{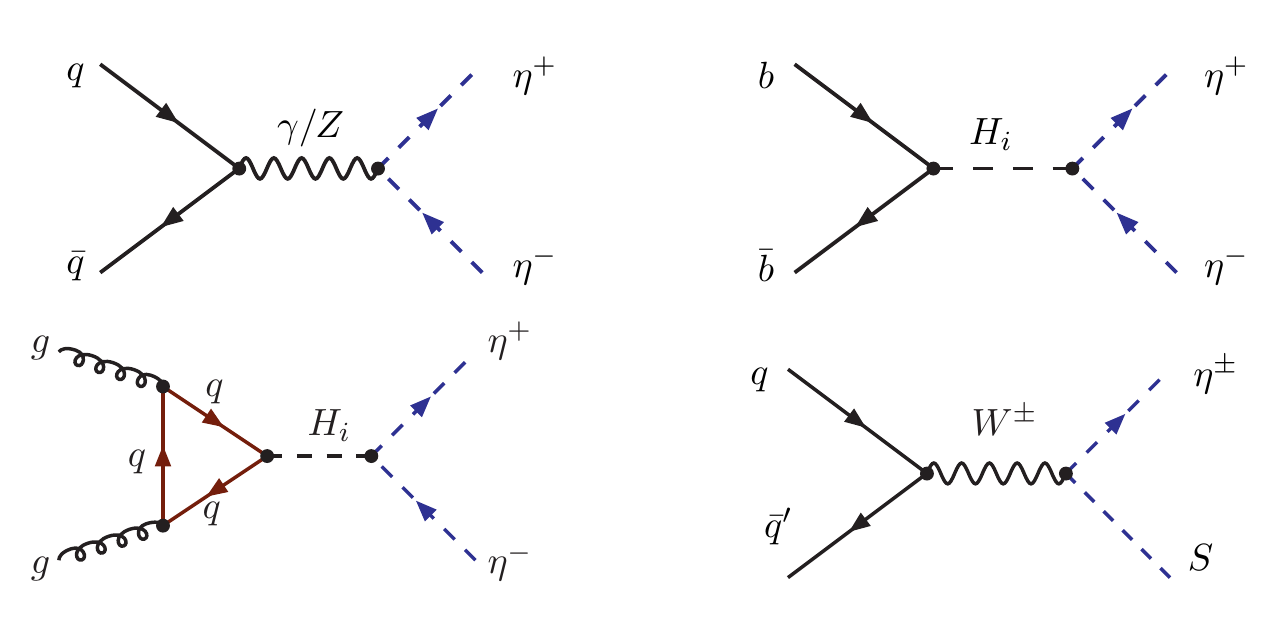}
\caption{Direct production channels
\label{fig:al}}
\end{center}
\end{figure}

For direct production at the parton level we have
\begin{subequations}
\label{parton-level}
\begin{equation}
\label{parton-level-1}
q{\bar q}^\prime\rightarrow W^{\pm\star} \rightarrow \eta^\pm S,
\end{equation}
\begin{equation}
\label{parton-level-2}
q{\bar q}\rightarrow \gamma^\star, Z^\star, H_i^\star \rightarrow \eta^+ \eta^-,
\end{equation}
\begin{equation}
\label{parton-level-3}
gg\rightarrow H_{i}^\star \rightarrow \eta^+ \eta^-, \qquad i=1,2,3.
\end{equation}
\end{subequations}
Some of these processes are shown in Fig.~\ref{fig:al}.
\subsection{Charged scalar decay}
\label{Subsect:charged_scalar_decay}
\begin{figure}[htb] 
\centering
\includegraphics[scale=0.6]{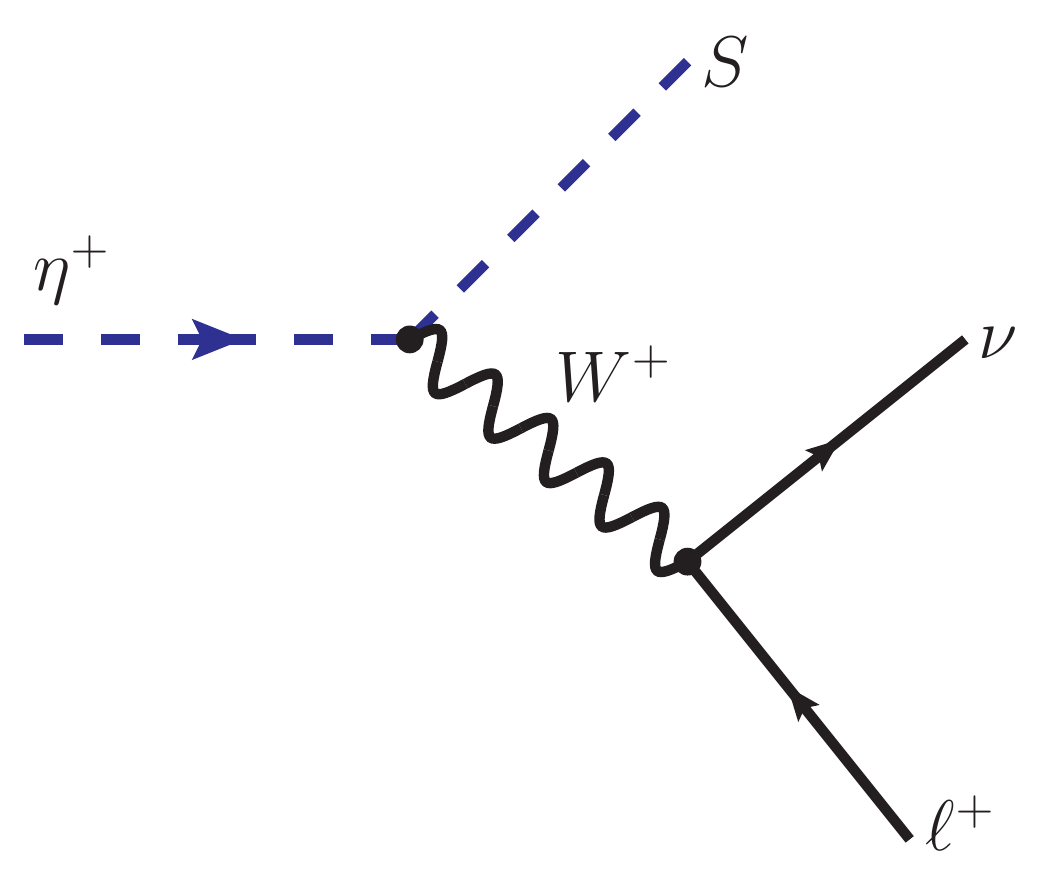}
\caption{Decay of a charged scalar $\eta^+$ to the DM particle $S$, a charged lepton and a neutrino). 
\label{Fig:eta-decay}}
\end{figure}

In favorable situations, decays of $\eta^\pm$ could lead to observable signals. Small mass splitting between the charged scalar, $\eta^\pm$, and the inert one,  $S$, can lead to long-lived charged scalars that give rise to displaced vertices in LHC detectors. (A special version of this scenario has been proposed in Ref.~\cite{Huitu:2010uc}.) In the case of a heavy $S$, the IDM2 requires a small splitting in order to give the correct DM abundance. In the case of a ``light'' dark matter ($M_S=75~\text{GeV}$) we find that in a considerable part of parameter space the mass difference between $\eta^\pm$ and $S$ could also be small. If charged scalars are produced, they can decay to $S$ accompanied by a $W^\pm$ (perhaps virtual) which then decays to fermions (as depicted in Fig.~\ref{Fig:eta-decay}). (Depending on the mass hierarchy between particles of the inert sector, also an $A$ could be an intermediate state \cite{Grzadkowski:2010au}.)

We shall here focus on the decay channels
\begin{equation}
\eta^+ \to S u \bar d,
\end{equation}
and
\begin{equation}
\eta^+ \to S \ell^+\nu_\ell,
\end{equation}
the latter being depicted in Fig.~\ref{Fig:eta-decay}. In the contact-interaction limit, the charged-lepton spectrum is given by
\begin{equation}
\frac{d\Gamma}{dE^+}
=\frac{\GF^2\Meta}{4\pi^3}
\frac{(\Meta^2-2\Meta E^+-M_S^2)^2}{(\Meta^2-2\Meta E^+)^2}
(\Meta-2E^+)(E^+)^2,
\end{equation}
where $E^+$ denotes the charged-lepton energy.

For small mass gaps
\begin{equation}
\Delta\equiv \Meta-M_S,
\end{equation}
the above expression simplifies,
\begin{equation}
\frac{d\Gamma}{dE^+}
=\frac{\GF^2\Meta}{\pi^3}
(\Delta-E^+)^2(E^+)^2,
\end{equation}
and the integrated width is
\begin{equation}
\Gamma(\eta^+\to S \ell^+\nu)
=\frac{G_\text{F}^2}{30\pi^3}(\Meta-M_S)^5.
\end{equation}

The experimental signature would be the observation of a charged $\eta^\pm$ track from the production point up to the decay vertex together with a kink corresponding to the decay of the charged scalar. Such a kink does not depend on the nature of the accompanying boson being produced; whether it is $\eta^\mp$, $A$ or $S$, at least one kink is always there. Missing energy through the presence of the two $S$'s in the final state will also be present. 

\begin{figure}[htb] 
\centering
\includegraphics[scale=0.7]{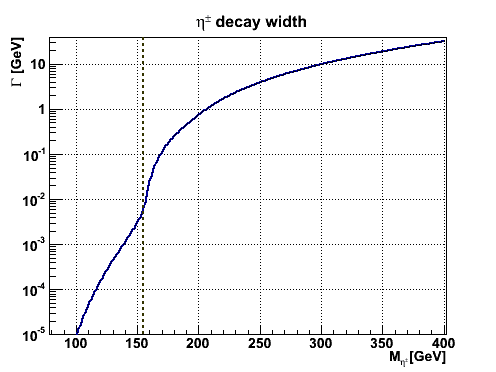}
\caption{Decay width of the $\eta^\pm$, for $M_S=75~\text{GeV}$ and $M_A=110~\text{GeV}$. Threshold for $\eta^\pm\to SW^\pm$ have been indicated.
\label{Fig:eta_width}}
\end{figure}

\begin{figure}[htb] 
\centering
\includegraphics[scale=0.7]{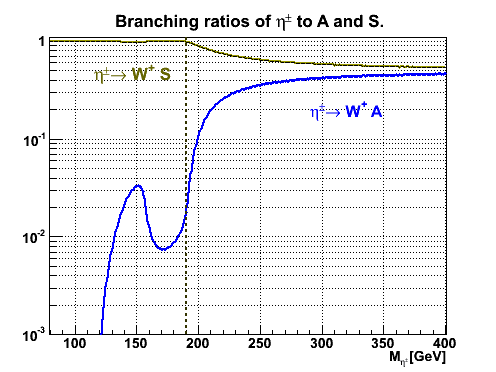}
\caption{$\eta^\pm$ branching ratios to $S$ or $A$, plus two fermions (via a $W$), for $M_S=75~\text{GeV}$ and $M_A=110~\text{GeV}$. The threshold for $\eta^\pm\to AW^\pm$ has been indicated.
\label{Fig:eta_br}}
\end{figure}

The $\eta^\pm$ width will be determined by its decays to the DM particle $S$ accompanied by two fermions, as well as a similar decay, if kinematically possible, to the pseudoscalar $A$ and two fermions,
\begin{equation}
\Gamma=\Gamma(\eta^\pm\to S W^\pm)+\Gamma(\eta^\pm\to A W^\pm),
\end{equation}
where the $W$ may be virtual. In fact, the width rises steeply as the $W$ reaches threshold, at $M_{\eta^\pm}=M_S+M_W$,
see Fig.~\ref{Fig:eta_width}. The $A$ would in turn decay, via a virtual or real $Z$ (there is no $ASH_j$ coupling \cite{Grzadkowski:2010au}), to the $S$ and two fermions. Such chains would thus yield four fermions, occasionally three of them would be charged leptons.

In Fig.~\ref{Fig:eta_br}, we show the branching ratios. If the $S$ is significantly lighter than the $A$, then the $Sf\bar f'$ final state would dominate. The oscillations (Fig.~\ref{Fig:eta_br}) in the $W^\pm A$ branching ratio are due to $WS$ reaching threshold for real $W$s, followed by $WA$ reaching threshold for real $W$s.

\section{Allowed parameter domains}
\label{Sec:allowed-parameters}
\setcounter{equation}{0}

The model has a few more parameters than the IDM, since it allows CP violation in the scalar sector. Compared with the 2HDM, the additional parameters are those related to the ``inert'' doublet. In Refs.~\cite{Grzadkowski:2009bt,Grzadkowski:2010au}, some parameters were set to ``interesting'' values, and the rest of them were scanned over in order to determine allowed domains. Since then, results from the LHC have imposed additional constraints on the model, as explored in Ref.~\cite{Basso:2012st} and summarized in section~\ref{Subsect:obs-constraints}.

\begin{figure}[htb]
\centering
\includegraphics[scale=0.55]{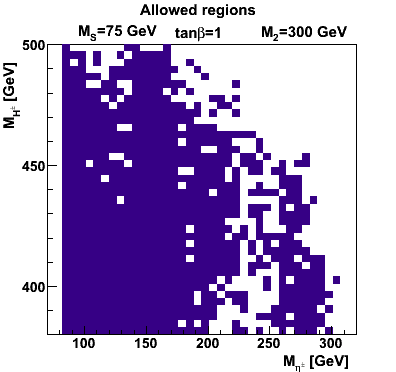}
\includegraphics[scale=0.55]{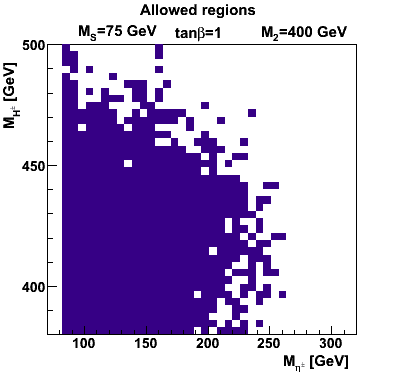}
\caption{Representitive allowed parameter region in the $M_{\eta^\pm}$--$M_{H^\pm}$ plane at tan$\beta=1$. Two values of $M_2$ are considered, 300 and 400~GeV.}
\label{Fig:allowed-mh_ch-m_eta_pm-tanb1}
\end{figure}

In this section, we assume that the following parameters are fixed:
 \begin{gather}
\text{Inert sector:} \quad
 M_{S}= 75~\text{GeV},\quad M_{A}= 110~\text{GeV}, \\
 \text{2HDM sector:} \quad
M_{1}= 125~\text{GeV},\quad 
\mu= 200~\text{GeV},
\end{gather}
where $\mu$ is the familiar (soft-) mass parameter of the 2HDM \cite{Grzadkowski:2009bt}.
The value for $M_S$ is taken in the central region of the allowed low-energy region of the IDM \cite{LopezHonorez:2006gr} and not excluded by the XENON100 data \cite{Aprile:2011hi,Aprile:2012nq,Grzadkowski:2011zn}, whereas the value for $M_A$ is taken as low as is compatible with the LEP results \cite{Lundstrom:2008ai}.
Furthermore,
$\tan\beta$ is chosen to be low (1 or 2) and  $M_{H^\pm}$ is taken to be between 380 and 500~GeV. The mass of the charged scalar, $M_{\eta^\pm}$, and the mass parameter of the inert potential, $m_\eta$, are scanned over, together with $\alpha_1,\alpha_2,\alpha_3$, the parameters determining mixing in the neutral-Higgs sector (see \cite{Grzadkowski:2009bt,Grzadkowski:2010au}). As soon as the scan reaches an acceptable point in the $\alpha$ space of the 2HDM sector it proceeds to the next point in the ``outer'' parameter space. We do not respect a particular mass hierarchy between $M_{\eta^{\pm}}$ and $M_A$, thus the scan over $M_{\eta^\pm}$ starts from 80~GeV and proceed upwards. 

\begin{figure}[htb]
\centering
\includegraphics[scale=0.55]{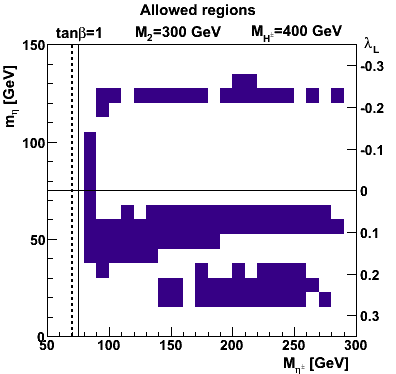}
\includegraphics[scale=0.55]{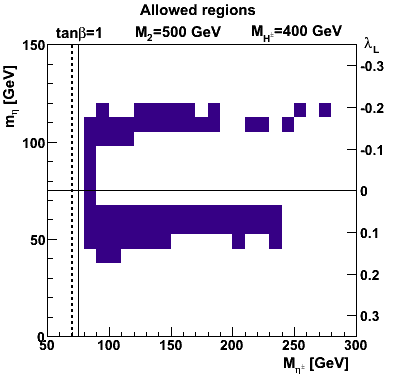}
\caption{Allowed parameter region (blue) at $\tan\beta=1$, $M_{H^\pm}=400~\text{GeV}$ and two values of $M_2$, 300 and 500~GeV.}
\label{Fig:allowed-mh_ch-m_eta_pm-tanb1-400}
\end{figure}

For $\tan\beta=1$ we show in Fig.~\ref{Fig:allowed-mh_ch-m_eta_pm-tanb1} allowed regions in the $M_{\eta^\pm}$--$M_{H^\pm}$ plane, for the two values $M_2=300~\text{GeV}$ and 400~GeV. For a given value of $M_{H^\pm}$, the cut-off at high $M_{\eta^\pm}$ is due to the contribution of the $\eta$ doublet to $T$, arising from the mass splitting $M_{\eta_\pm}-M_S$ and $M_{\eta_\pm}-M_A$.
We recall that the splittings between the neutral ones ($A$ and $S$)
and the splitting between a neutral ($A$ or $S$) and the charged one
($\eta^\pm$) contribute to $T$ with opposite signs \cite{Grimus:2007if}.
Some of the ``empty'' points are clearly due to the finite number of points scanned over.

Turning now to a more detailed discussion, we consider in Fig.~\ref{Fig:allowed-mh_ch-m_eta_pm-tanb1-400} the two values $M_2=300~\text{GeV}$ and 500~GeV, and a fixed value of $M_{H^\pm}=400~\text{GeV}$. Here, we display allowed regions in the $M_{\eta^\pm}$--$m_\eta$ plane, where $m_\eta$ is a ``soft'' parameter  (\ref{v3}) related to the coupling $\lambda_L$ among the Higgs and the inert doublet by Eq.~(\ref{Eq:lambda_L}). 

\begin{figure}[htb]
\centering
\includegraphics[scale=0.55]{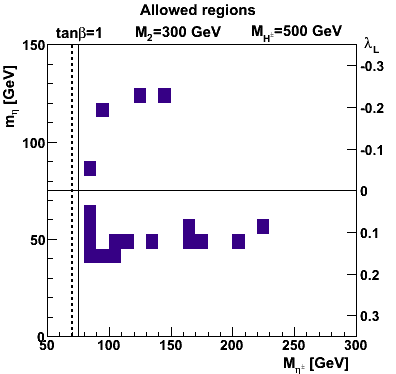}
\includegraphics[scale=0.55]{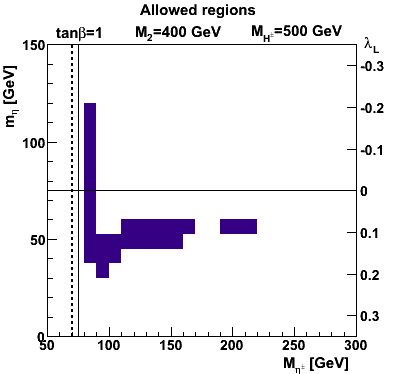}
\caption{Similar to Fig.~\ref{Fig:allowed-mh_ch-m_eta_pm-tanb1-400}, for $M_{H^\pm}=400~\text{GeV}$, with $M_2=300~\text{GeV}$ and 400~GeV.}
\label{Fig:allowed-mh_ch-m_eta_pm-tanb1-500}
\end{figure}

\begin{figure}[htb]
\centering
\includegraphics[scale=0.55]{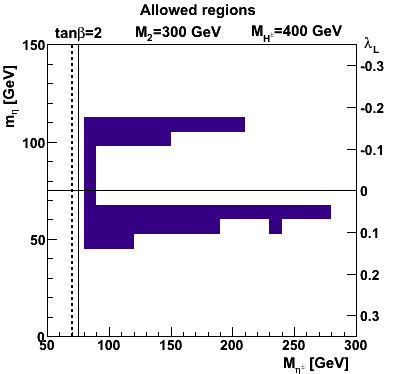}
\includegraphics[scale=0.55]{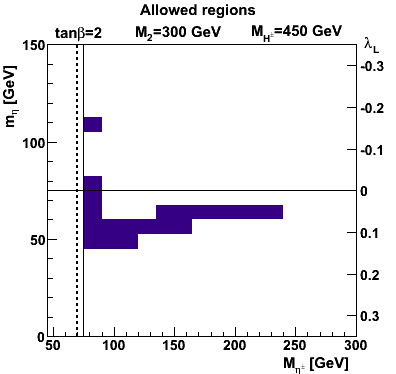}
\caption{Representative allowed parameter region at small tan$\beta$.}
\label{Fig:allowed-mh_ch-m_eta_pm-tanb2}
\end{figure}

At low values of $M_{\eta^\pm}$ (around 100~GeV), a range of $m_\eta$ (or $\lambda_L$) values are acceptable, as the $S$ and $\eta^\pm$ could in the early universe have co-annihilated via a virtual $W$ \cite{Grzadkowski:2010au}, thereby leading to an acceptable DM amount. At higher masses ($M_{\eta^\pm}$), this co-annihilation mode is less efficient since the $W$ will be further off-shell, so a minimal coupling $\lambda_L$ to the Higgs sector is required, in order to avoid overproduction of dark matter in the early universe. Thus, a band around $\lambda_L=0$ is excluded. At large values of $|\lambda_L|$, there is a cut-off, as such parameters would result in too much annihilation in the early universe via the $SS\to H_i$ channels. Again, some of the ``empty'' points are due to the finite number of points in the scans.

Similarly, we consider in Fig.~\ref{Fig:allowed-mh_ch-m_eta_pm-tanb1-500} the case of $M_{H^\pm}=500~\text{GeV}$ and the two values $M_2=300~\text{GeV}$ and 400~GeV. In this case, a much smaller part of the plane is allowed. With $\tan\beta=2$ (Fig.~\ref{Fig:allowed-mh_ch-m_eta_pm-tanb2}), less parameter space is allowed than for $\tan\beta=1$. As discussed above, the cut-off at high values of $M_{H^\pm}$ is basically due to the electroweak precision data, via the constraint on $T$.

\subsection{Benchmark points}

For the purpose of studying representative production cross sections in Sect.~\ref{Sec:LHC}, we shall in the following consider some selected benchmark points in the parameter space which pass all the aforementioned theoretical and experimental constraints, given in Table~\ref{tab:points}. We note that the 2HDM benchmark points of Ref.~\cite{Basso:2012st} are not automatically allowed, since the presence of the charged $\eta^\pm$ field will modify the $H_1\to\gamma\gamma$ branching ratio. Likewise, there will be invisible decays of $H_2$ and $H_3$ to $SS$ and $AA$, as well as $H_{2,3}\to \eta^+\eta^-$ if kinematically allowed.

\begin{table}[ht]
\begin{center}
\begin{tabular}{|c|c|c|c|c|c|c|c|c|c|c|}
\hline
\ & $\alpha_1/\pi$ & $\alpha_2/\pi$ & $\alpha_3/\pi$ & $\tan\beta$ & $M_2$ & $M_3$ & $M_{H^\pm}$
& $R_{\gamma\gamma}^\text{2HDM}$ & $R_{\gamma\gamma}^\text{IDM2}$ & $R_{\gamma\gamma}^\text{IDM2}$\\
\hline
 $P_1$ & $0.39$ & $-0.026$ & $0.46$ & $1$ & $300$ & 333 & 400 & 0.92 & 0.50 & 0.34\\
 $P_2$ & $0.39$ & $-0.009$ & $0.025$  & $1$ & $300$ & 325 & 400 & 0.91 & 0.50 & 0.34 \\
 $P_3$ & $0.37$ & $-0.018$ & $0.016$ & $1$ & $300$ & 394 & 400 & 0.78 & 0.45 & 0.30 \\
 $P_4$ & $0.45$ & $-0.19$ & $0.38$ & $1$ & $300$ & 443 & 400 & 1.57 & 0.77 & 0.63 \\
 $P_5$ & $0.15$ & $0.42$ & $0.45$ & $1$ & $300$ & 504 & 400 & 3.56 & 1.69 & 1.63 \\
 $P_6$ & $0.40$ & $-0.04$ & $0.07$ & $1$ & $300$ & 348 & 400 & 0.99 & 0.54 & 0.37  \\
\hline
$P_7$ & $0.28$ &  $-0.46$ & $0.11$ & $1$ & $400$ & 467 & 450 & 4.02 & 1.84 & 1.80 \\
$P_8$ & $0.48$ &  $-0.11$ & $0.31$ & $1$ & $400$ & 518 & 450 & 0.99 & 0.50 & 0.36 \\
\hline
$P_9$ & $0.20$ &  $0.45$ & $0.38$ & $1$ & $400$ & 487 & 500 & 3.71 & 1.71 & 1.67 \\
\hline
$P_{10}$ & $0.16$ &  $-0.44$ & $0.49$ & $2$ & $300$ & 468 & 400 & 0.26 & 0.52 & 0.50 \\
$P_{11}$ & $0.45$ &  $-0.01$ & $0.39$ & $2$ & $300$ & 325 & 400 & 0.84 & 0.64 & 0.43 \\
\hline
\end{tabular}
\end{center}
\caption{Benchmark points selected from the allowed 2HDM parameter space. Masses are in GeV, $\mu=200~\text{GeV}$.  Values of the ratio $R_{\gamma\gamma}$ are given for the 2HDM, as well as for the present model. Two values of $M_{\eta^\pm}$ are considered, 100~GeV (next-to-last column) and 200~GeV (last column).
\label{tab:points}}
\end{table}

The LHC experiments  \cite{atlas:2012gk,cms:2012gu} indicate an $H_1\to\gamma\gamma$ signal somewhat higher than that of the Standard Model. We quote in Table~\ref{tab:points} the value of this branching ratio, relative to that of the SM,
\begin{equation} \label{Eq:R_gammagamma}
R_{\gamma\gamma}=\frac{\Gamma(H_1\to gg){\rm BR}(H_1\to\gamma\gamma)}
{\Gamma(H_\text{SM}\to gg){\rm BR}(H_\text{SM}\to\gamma\gamma)}.
\end{equation}
This ratio is actually quite a bit lowered by the contribution of the $\eta^\pm$ in the loop. 
Here, we have taken $M_{\eta^\pm}=100~\text{GeV}$ (next-to-last column) and $M_{\eta^\pm}=200~\text{GeV}$ (last column).
For comparison, its value in the absence of the $\eta^\pm$ contribution (i.e., for the 2HDM with the same parameters), is also given.

\section{Charged scalar production at the LHC}
\label{Sec:LHC}
\setcounter{equation}{0}

Because of the unbroken $Z_2$ symmetry associated with the $\eta$ doublet, members of the doublet can only be pair produced.
Charged scalars can thus either be pair produced, or singly produced, but then in association with a neutral member of the doublet, $S$ or $A$. We shall in the following consider the two processes
\begin{equation}
pp\to\eta^\pm S X,
\end{equation}
and
\begin{equation}
\label{Eq:sigma-eta-eta}
pp\to\eta^+\eta^- X.
\end{equation}
Members of the $\eta$ doublet do not couple directly to quarks or gluons, but can be pair produced via a photon (Drell--Yan mechanism), a $Z$, or a Higgs boson $H_i$. 

The model has been implemented through the LanHEP module \cite{Semenov:2010qt} (see \cite{Mader:2012pm} for details) and the following analysis has been performed by means of the CalcHEP package \cite{Belyaev:2012qa}. Furthermore, we have used the CTEQ6.6M \cite{Nadolsky:2008zw} set of five-flavour parton distribution functions (PDFs). Due to their relevance at hadron colliders, the effective $ggH_i$ and $\gamma\gamma H_i$ vertices have been implemented by means of a link between CalcHEP and LoopTools \cite{Hahn:1998yk}.

\begin{figure}[htb]
 \begin{center}
\includegraphics[scale=0.38]{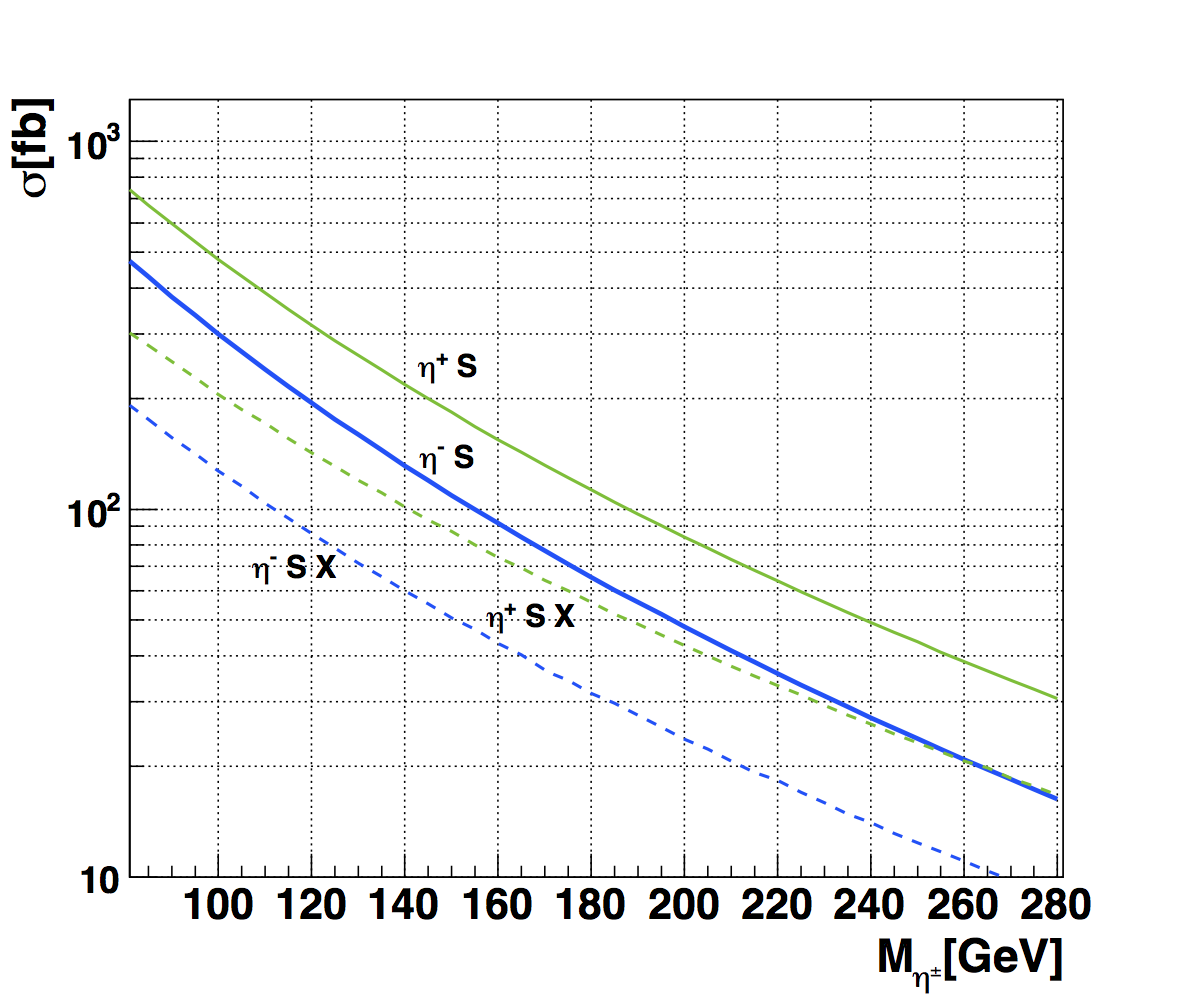}
\includegraphics[scale=0.38]{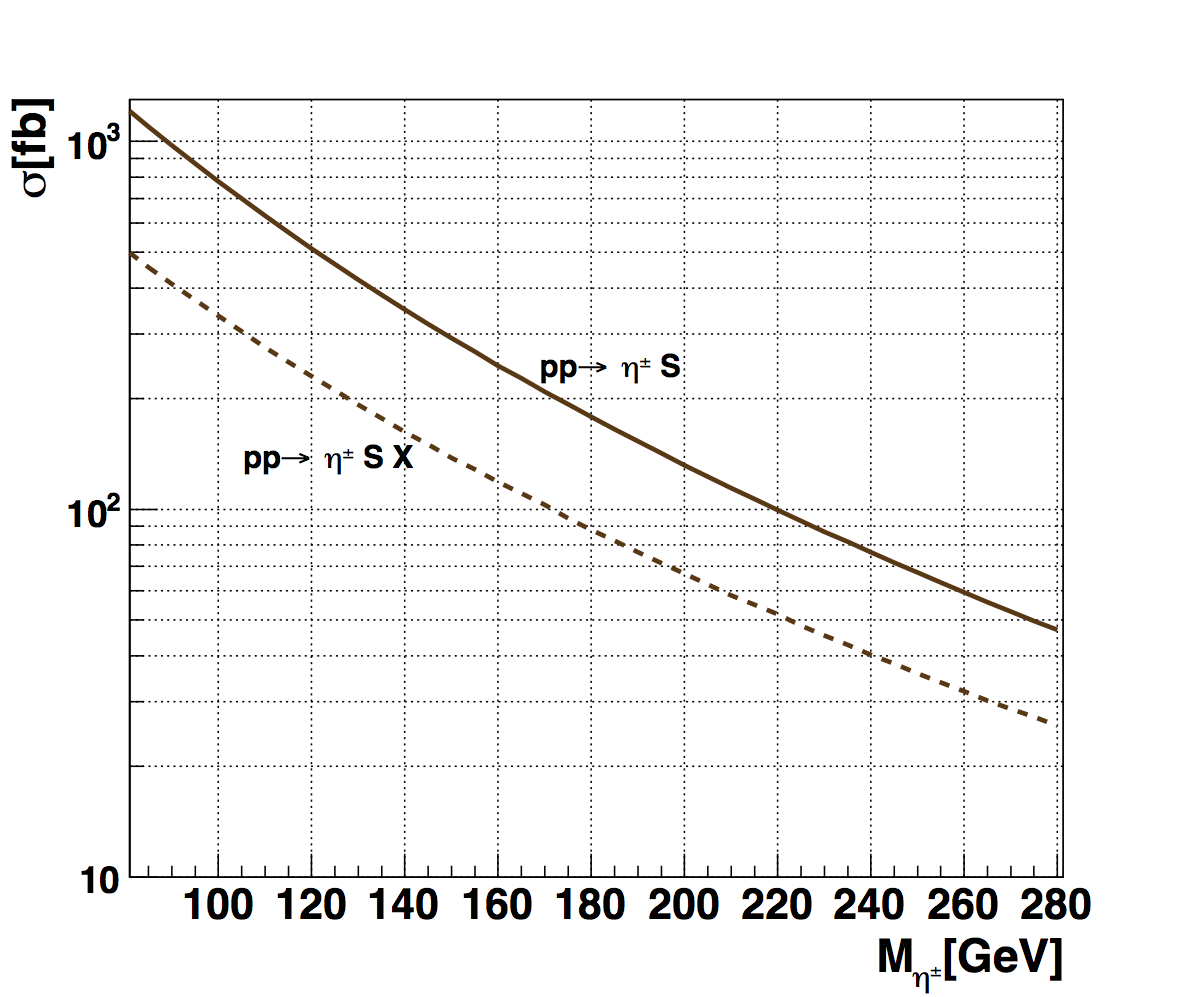}
\end{center}
\caption{Left: Individual cross sections for $\eta^+ S$ and $\eta^- S$, relevant for leptonic decay of $\eta^\pm$. Right: Sum, relevant for hadronic decay of the $W$ coming from $\eta^\pm\to SW^\pm$. In both panels, we consider $\sqrt{s}=14~\text{TeV}$, and the benchmark point $P_1$. The dashed curves refer to the case of an extra jet satisfying the cuts of Eq.~(\ref{Eq:cuts}).}
\label{Fig:sigma-eta-S}
\end{figure}

\subsection{Associated production of $\eta^+ S$ and $\eta^- S$}
The associated production of a charged $\eta^\pm$ in association with the DM particle $S$ mainly proceeds via an $s$-channel $W$-exchange. Since there are more $u$ quarks than $d$ quarks in the protons, the rate for 
\begin{equation}
u \bar d \to W^+ \to \eta^+ S
\end{equation}
will be higher than that for
\begin{equation}
d \bar u \to W^- \to \eta^- S.
\end{equation}
Thus, there will be more $\eta^+$ produced than $\eta^-$.
We show these cross sections separately in Fig.~\ref{Fig:sigma-eta-S} for $P_1$.
When the virtual $W^\pm$ from the $\eta^\pm\to W^\pm S$ transition converts to jets, this difference is immaterial. However, when the $W^\pm$ decays leptonically, the lepton charge is of interest. The summed cross section is of the order of 100--1000~fb, out to a mass $M_{\eta^\pm}$ of the order of 150~GeV.

Another interesting observable is represented by the the previous final state with the addition of a hard jet, since this could trigger a specific experimental detection. In this instance, the cross section gets reduced as we can infer from the dashed curves in Fig.~\ref{Fig:sigma-eta-S}, in which the following kinematic cuts have been applied:
\begin{gather}
p_\text{T}^{\min}=20~\text{GeV}, \nonumber \\
-4.5<\eta(\text{jet})<4.5.
\label{Eq:cuts}
\end{gather}

\begin{figure}[htb]
 \begin{center}
\includegraphics[scale=0.27]{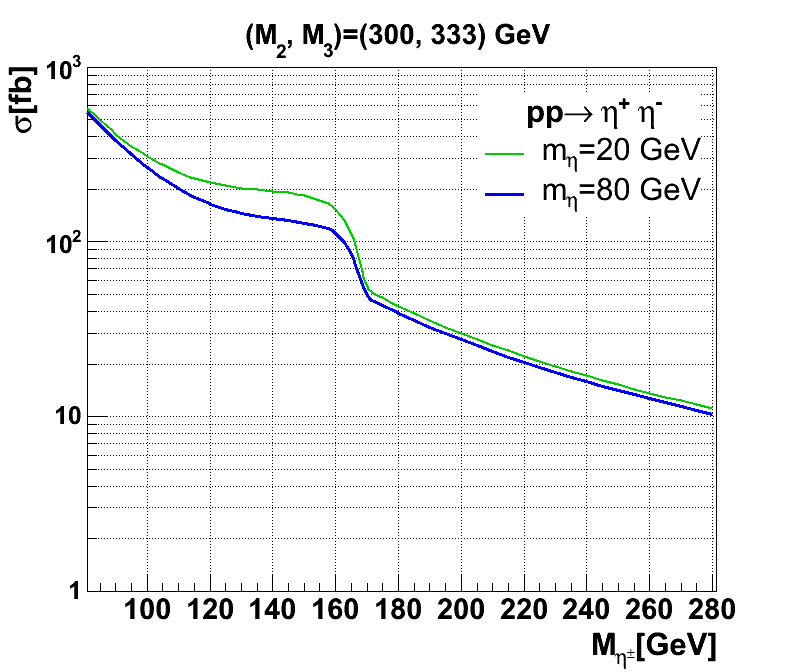}
\includegraphics[scale=0.27]{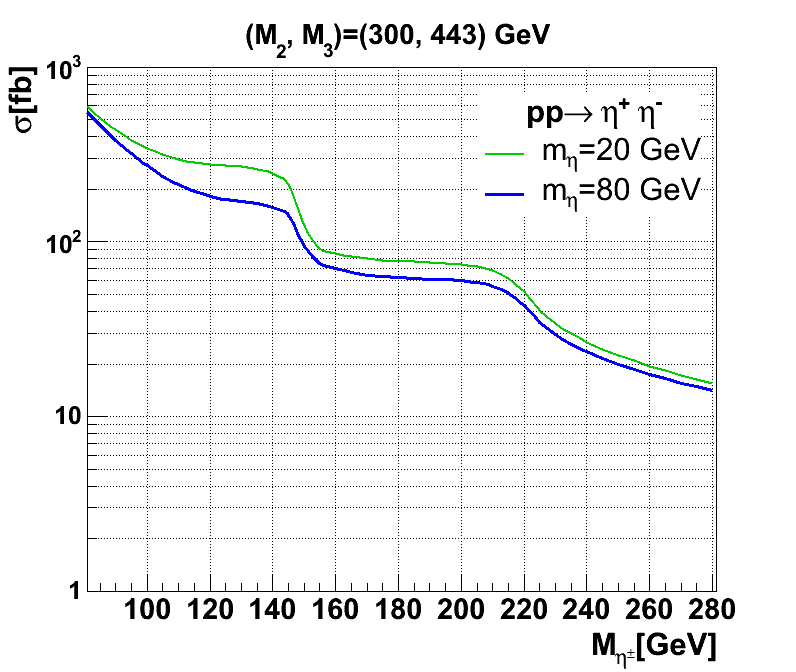}
\includegraphics[scale=0.27]{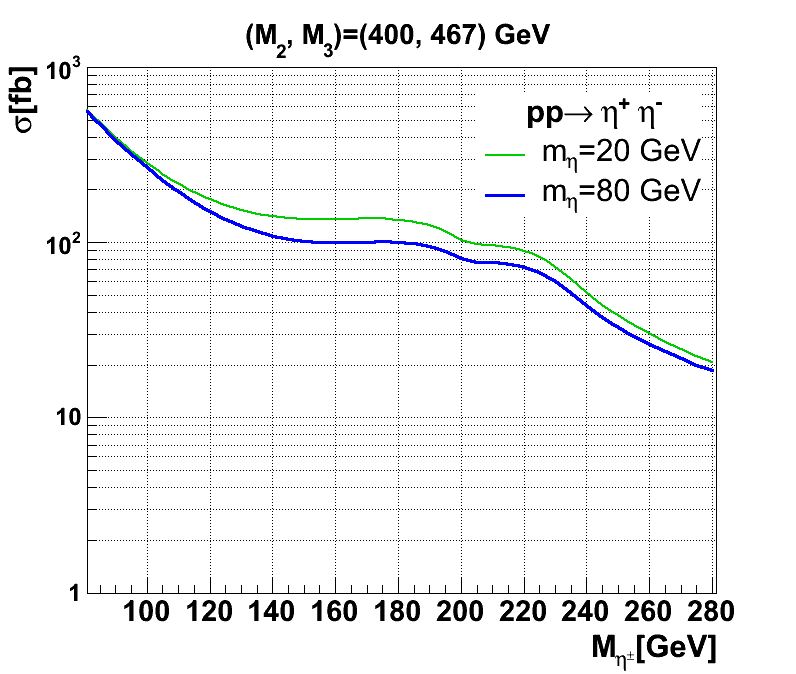}
\includegraphics[scale=0.27]{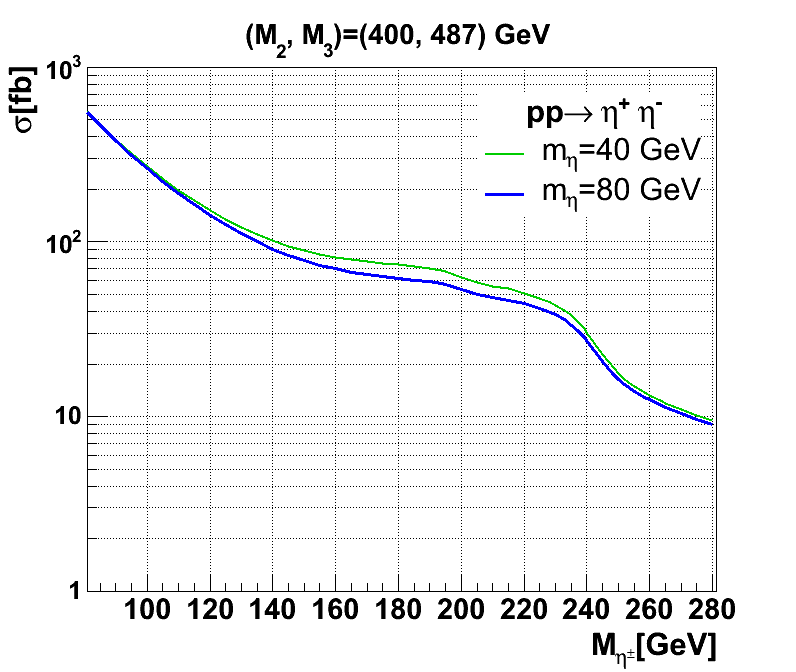}
\end{center}
\caption{Cross sections for $\eta^+\eta^-$ pair production at $\sqrt{s}=14~\text{TeV}$.
Top left: $P_1$ with $(M_2,M_3,M_{H^\pm})=(300,333,400)~\text{GeV}$, Top right: $P_4$ with $(M_2,M_3,M_{H^\pm})=(300,443,400)~\text{GeV}$.
Bottom left: $P_7$ with $(M_2,M_3,M_{H^\pm})=(400,467,450)~\text{GeV}$, Bottom right: $P_{9}$ with $(M_2,M_3,M_{H^\pm})=(400,487,500)~\text{GeV}$.}
\label{Fig:sigma-eta-eta}
\end{figure}

\subsection{Pair production of two charged particles, $\eta^+\eta^-$}

We present in Fig.~\ref{Fig:sigma-eta-eta} cross sections for the process (\ref{Eq:sigma-eta-eta}), for the benchmark points $P_1$ and $P_4$, with $(M_2, M_{H^\pm})=(300,400)~\text{GeV}$,  for $P_7$, with $(M_2, M_{H^\pm})=(400,450)~\text{GeV}$ as well as $P_{9}$, with $(M_2, M_{H^\pm})=(400,500)~\text{GeV}$. The shoulders observed are obviously due to the cut-offs from contributions involving $s$-channel $H_2$ and $H_3$ exchange. The cross section has some dependence on the ``soft'' parameter $m_\eta$, for which two values are considered. We recall that the $H_i\eta^+\eta^-$ coupling is given by
\begin{equation}
\eta^+\eta^-H_j: \quad -i\lambda_a vF_j, 
\end{equation}
with
\begin{equation}
F_{j}=\cos\beta R_{j1}+\sin\beta R_{j2},
\end{equation}
and $\lambda_a$ related to $m_\eta$ according to Eq.~(\ref{Eq:lambda-vs-splitting}).
Furthermore, $R$ is the neutral-sector mixing matrix \cite{Grzadkowski:2009bt,Grzadkowski:2010au}.
Thus, the two different values of $m_\eta$ correspond to two different strengths of the $H_i\eta^+\eta^-$ coupling.
The shoulders seen in Fig.~\ref{Fig:sigma-eta-eta} correspond to $2M_{\eta^\pm}=M_2$ and $2M_{\eta^\pm}=M_3$, for various values of these masses. There is of course no such structure corresponding to $H_1$ exchange in the $s$-channel, since this will be far off-shell for the masses considered, $M_1<M_{\eta^\pm}+M_S$.
The cross section is of the order of 100--500~fb, out to a mass $M_{\eta^\pm}$ of the order of 150~GeV.

\section{Experimental Possibilities}
\label{Sec:experimental}
\setcounter{equation}{0}

Far from the intention of carrying out a complete signal-over-background analysis, we devote the present section to the profiling of the $\eta^\pm$ signature at the LHC. Starting from the cross-section study performed in the previous section, we assume that the high-energy/high-luminosity run at the LHC is the only realistic chance to detect an emerging signal. 
Among the possibilities sketched in Eq.~(\ref{eq:prod}), one channel is available only at hadronic colliders, namely the single production. This occurs because it is triggered by an intermediate $W$, which can only be produced in hadron-hadron scatterings. For this, we focus on the simplest $\eta^\pm$ production channel, i.e. $p+p\to \eta^\pm+S/A$, at the centre-of-mass energy $\sqrt{s}=14$ TeV and the integrated luminosity $\Lumint=100$ fb$^{-1}$. Furthermore, we recall that a real $\eta^\pm$ can decay through two allowed modes, which are $\eta^\pm \to W^\pm+S$ and $\eta^\pm \to W^\pm+A$ with an off-shell $W^\pm$ produced in association with a neutral (pseudo)scalar. However, we have seen in section~\ref{Sec:allowed-parameters} that the allowed parameter space is accessible only if $M_A>M_S$, hence $A$ could not be part of a stable final state. This brings us to the specific study of the channel
\begin{eqnarray}
pp\to \eta^\pm S\to W^\pm SS \qquad \mbox{($W^\pm$ is off-shell)}.
\end{eqnarray}

The experimental possibilities to discover such a charged state depend on the parameter space which rules the kinematics of the signal (i.e, on the ``dark sector'' spectrum), they don't directly depend on the parameters of the non-inert part of the potential. In fact, the coupling is given by the three-point vertex \cite{Grzadkowski:2010au}
\begin{equation}
S\eta^\pm W^\mp:\quad \frac{e}{(2\sin{\theta_W})}(p_S-p_\eta)^\mu,
\end{equation}
where the momenta flow inwards and the notation is self-explanatory. This implies that the following analysis will be valid for any of the benchmark points in Table~\ref{tab:points}\footnote{We remark that this statement is valid only for the single-$\eta^\pm$ production. The $\eta^+\eta^-$ production is affected by specific choices of the parameters of the potential, in particular the coupling to the non-inert neutral Higgses, which constitute the main portal for that process.}.

The final states can be classified in two sets, depending on the $W^\pm$ decay mode: a hadronic one (with $W\to 2j$) and a leptonic one (with $W\to l+\nu_l$). Due to the enormous SM single-lepton background, only the former channel has any chance to be detected. However, in certain kinematic circumstances, both channels are open to the possibility of an interaction-vertex displacement, as we have already discussed in sub\-section~\ref{Subsect:charged_scalar_decay}, and this would make them accessible to experimental detection. In the following subsections, we will consider all of these features and we will perform a substantial profiling of the signature stemming from the single $\eta^\pm$ produced at the LHC.

\subsection{Hadronic final state: jets plus missing energy}

The study of the hadronic final states in single $\eta^\pm$ production is connected with the IDM2 signature
\begin{eqnarray}
p+p\to 2j+2S \simeq 2j+\text{MET},
\end{eqnarray}
since $S$ is a stable inert particle which can only be revealed by a detector as transverse missing energy.

The analysis of this channel at the LHC can be related to the ATLAS mono-jet plus missing transverse momentum searches \cite{ATLAS-CONF-2012-147}. For this, we will apply a set of standard kinematic cuts adopted by the ATLAS experiment, i.e.,
\begin{eqnarray}\label{ATLAScuts}
p^T_j& > & 20 \mbox{ GeV},\nonumber \\
\left|\eta _j\right| & < & 4.5, \\
\Delta R_{jj} & > & 0.5 \qquad \mbox{(or jet merging applied)}; \nonumber
\end{eqnarray}
where $p^T_j$ represents each jet's transverse momentum, $\eta$ is its pseudo-rapidity and $\Delta R=\sqrt{(\Delta\eta)^2+(\Delta\phi)^2}$. Starting from the analysis of section~\ref{Sec:allowed-parameters}, we establish a set of benchmarks that involves several values of $M_S$ and $M_{\eta^\pm}$. In order to test several degrees of mass splitting, we chose to assume three different values for $M_S$, 70, 80 and 90~GeV, and to combine them with four possible values for $M_{\eta^\pm}$, namely 100, 120, 140 and 160~GeV. Therefore, we have performed an event-generation analysis by producing several n-tuples consisting of $10^5$ events for the process
\begin{eqnarray}
pp\to \eta^\pm S \to W^\pm SS,
\end{eqnarray}
and letting the off-shell $W^\pm$ decay inclusively. Hence, we have weighted each of them by its corresponding production cross section (we summarise the values in Table~\ref{tab:cs}). From these values we extract the information about the relatively simple behaviour of the cross section with respect to the final state masses: it is basically determined by the phase space, when masses increase then cross sections decrease.

\begin{table}[ht]
\begin{center}
\begin{tabular}{|c|c|c|c|}
\hline
& $M_S=70$ GeV & $M_S=80$ GeV  & $M_S=90$ GeV \\
\hline
 $M_{\eta^\pm}=100$ GeV & $708$ & $583$ & $483$ \\
 $M_{\eta^\pm}=120$ GeV & $463$ & $392$ & $333$ \\
 $M_{\eta^\pm}=140$ GeV & $310$ & $265$ & $219$ \\
 $M_{\eta^\pm}=160$ GeV & $222$ & $187$ & $149$ \\
\hline
\end{tabular}
\end{center}
\caption{Total cross section (in fb) for the process $pp\to W^\pm SS$ at the LHC with a centre-of-mass energy of $\sqrt{s}=14$ TeV. The kinematic cuts given by Eq.~(\ref{ATLAScuts}) are applied.
\label{tab:cs}}
\end{table}

After the generation is performed, we focus on the hadronic final state and apply the cuts in Eq.~(\ref{ATLAScuts}). After the $\Delta{R}$ criterion is applied, we obtain the first noticeable result: there is no surviving di-jet signal because all the di-jets are merged in an ``effective'' mono-jet. Apart from the $M_{\eta^\pm}=160$ GeV case, this occurs because the two jets stem from an off-shell $W^\pm$ and they have an invariant mass that is mostly related to the off-shell-ness. As a consequence, they release most of their energy via their common longitudinal momentum, so they are not separately observable by the detector. The only exception occurs when the $W^\pm$ is allowed to be on-shell: in that case we count a very small (not significant) fraction of final states that behave as two separate jets. In a nutshell, we are mostly dealing with an effective mono-jet plus MET signal, i.e.
\begin{eqnarray}
pp\to j+\text{MET}.
\end{eqnarray}

\begin{figure}[htb]
 \begin{center}
\includegraphics[scale=0.44]{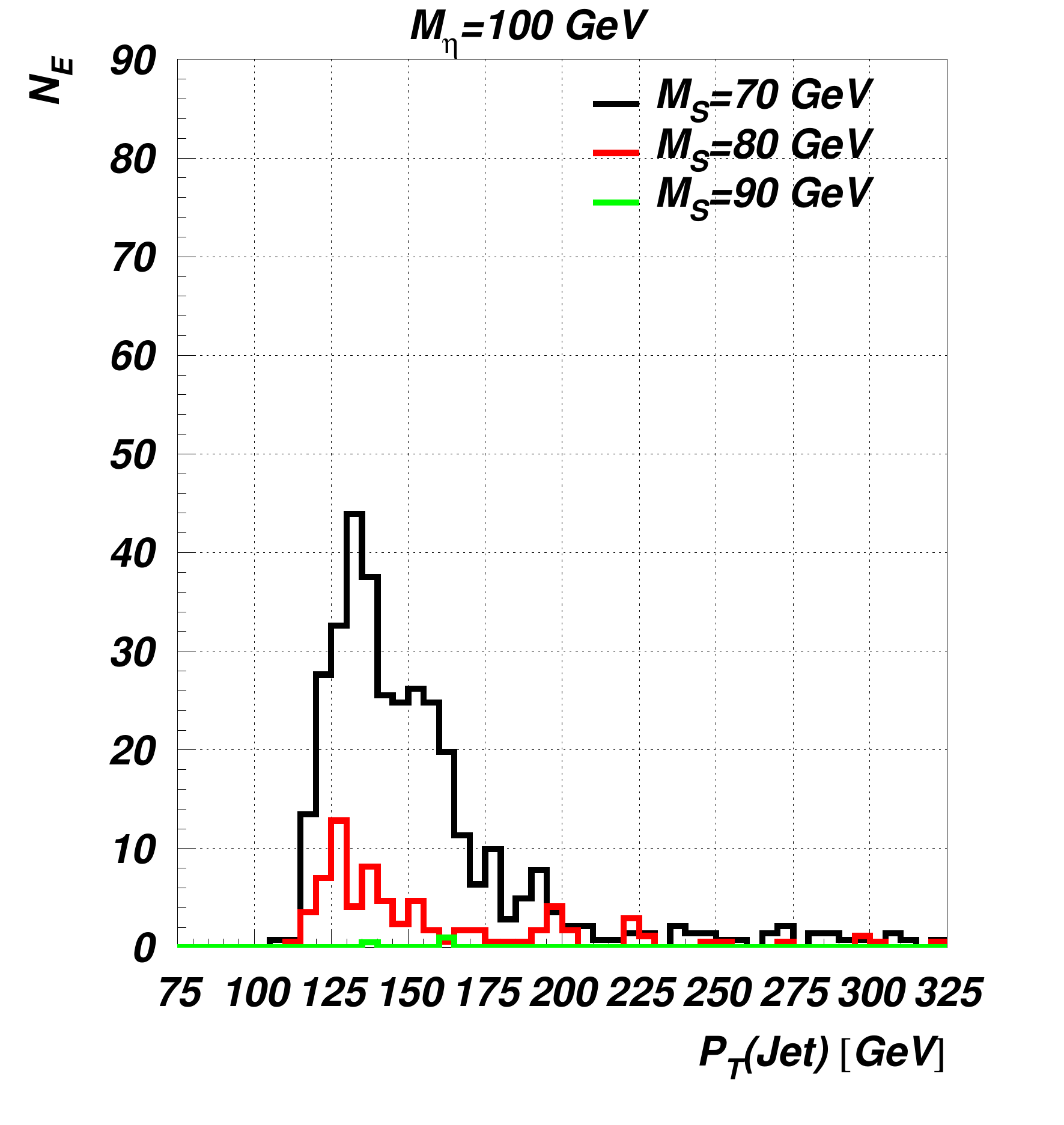}
\includegraphics[scale=0.44]{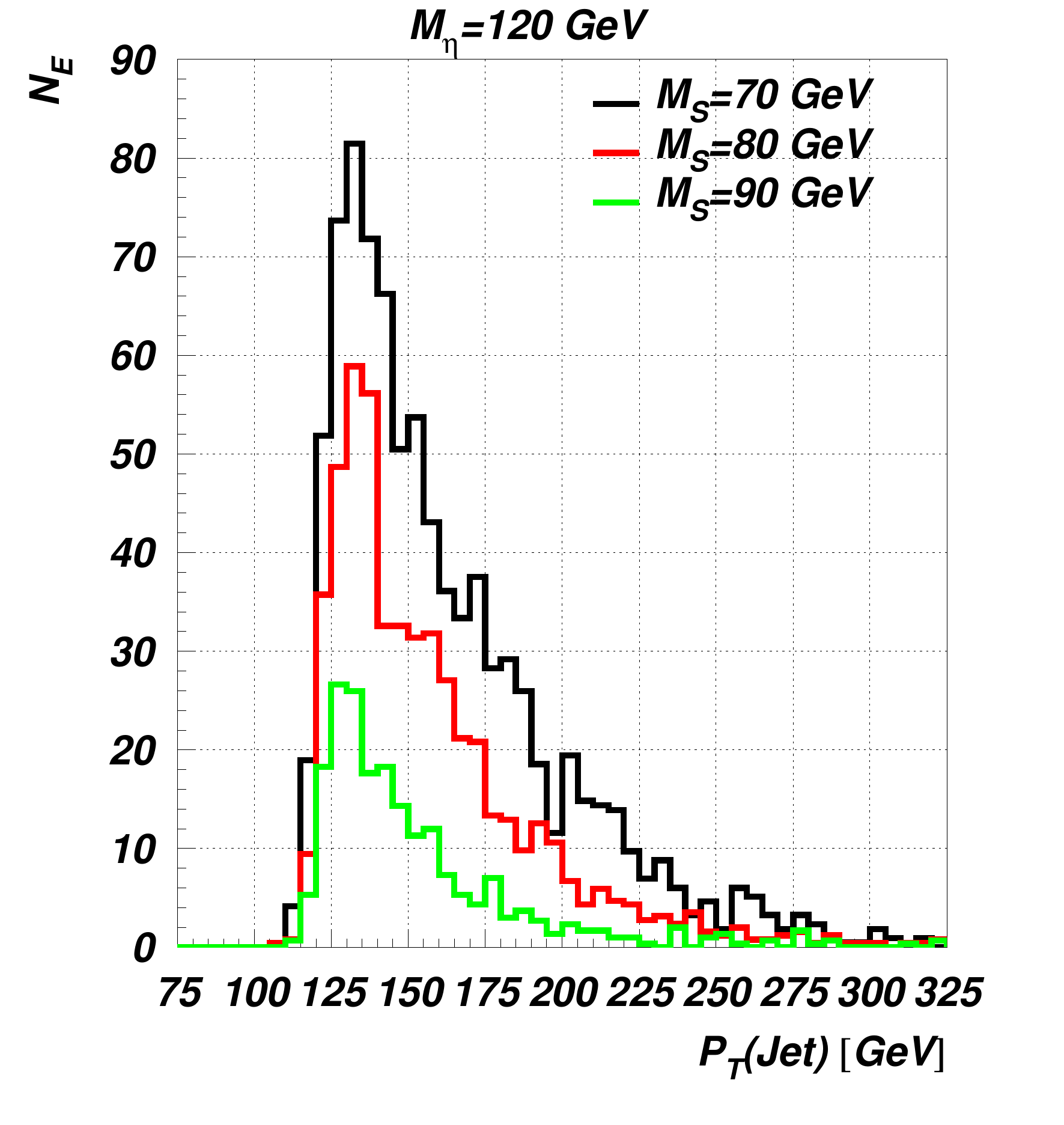} \\
\includegraphics[scale=0.44]{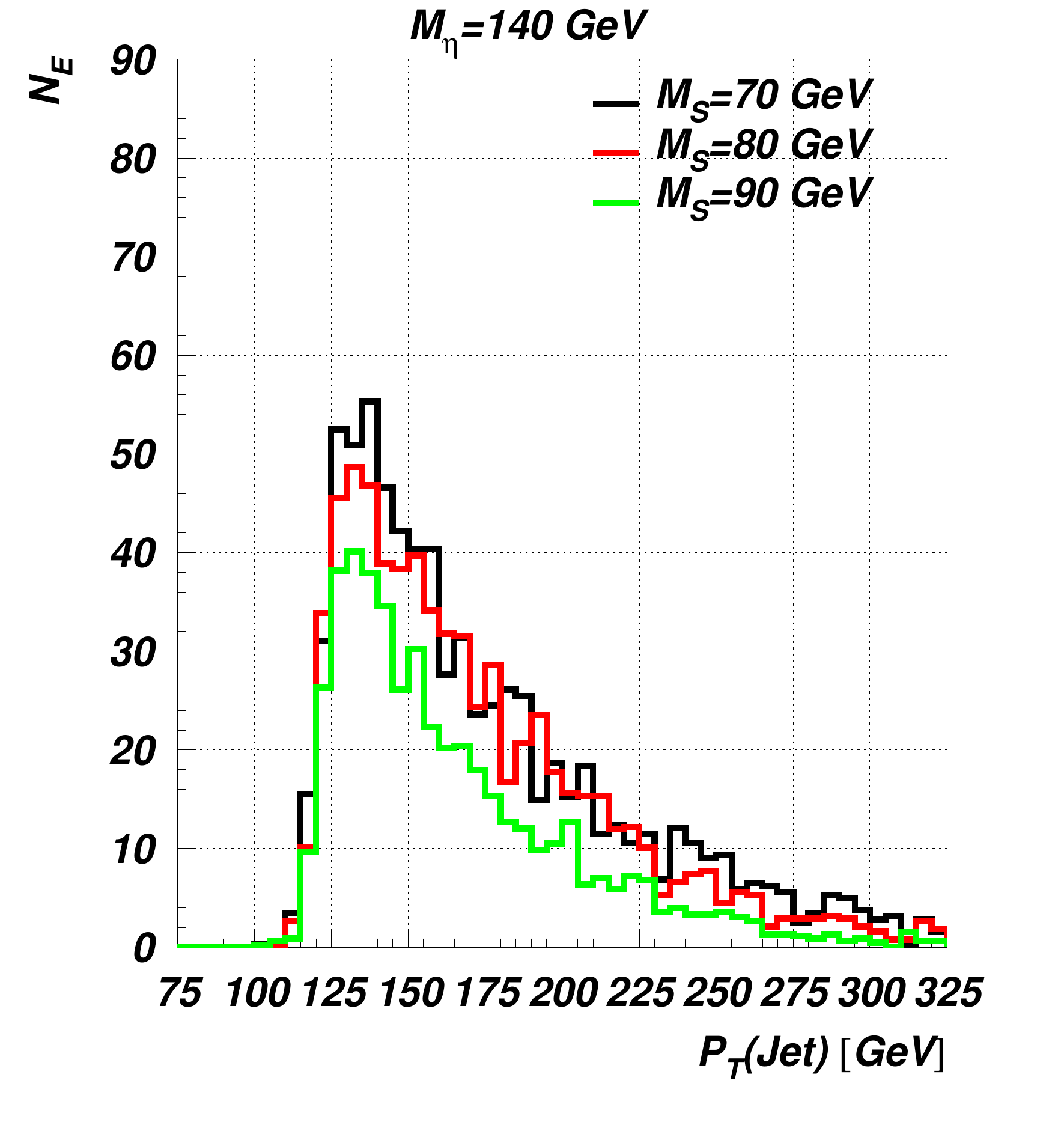}
\includegraphics[scale=0.44]{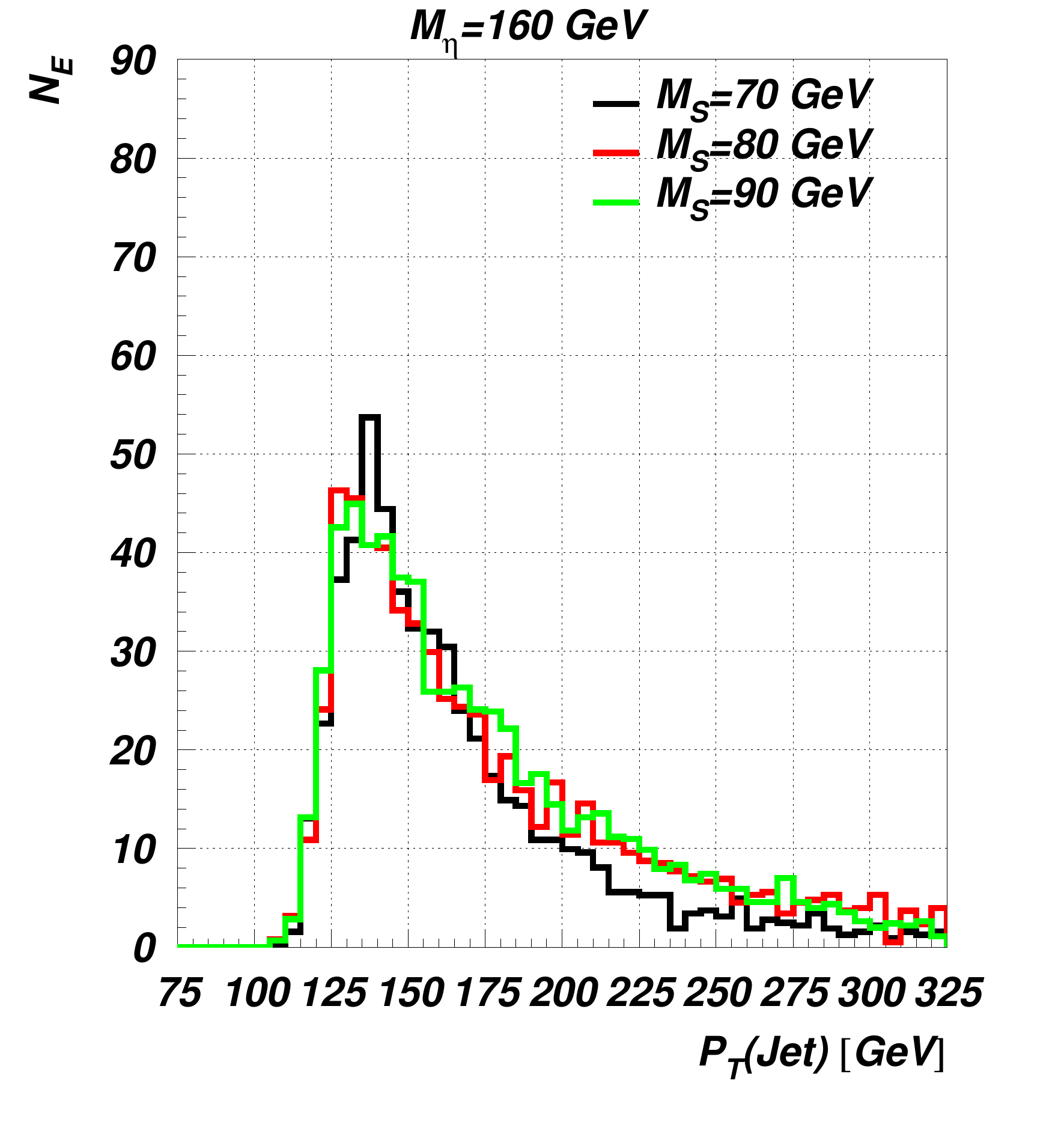}
\end{center}
\caption{Number of integrated events against the $p^T_j$ for the process $pp\to J+\text{MET}$ at the LHC with $\sqrt{s}=14$ TeV and $\Lumint=100$ fb$^{-1}$.
\label{Fig:graph_ptj}}
\end{figure}

Despite the fact that a study of the cutting strategy concerning the LHC high-energy scenario is surely premature at this stage, we mimic the analysis of the $\sqrt{s}=8$ TeV scenario by further requiring two selection criteria for the detection of the mono-jet signal which are stated in \cite{ATLAS-CONF-2012-147}, i.e.
\begin{equation}
\text{MET}>120~\text{GeV}, \qquad p^T_j>120~\text{GeV}.
\end{equation}
Then, we plot the number of events against the $p^T_j$ in $5$ GeV bins. The result is shown in Fig.~\ref{Fig:graph_ptj}.

From the frames of Fig.~\ref{Fig:graph_ptj} it is clear that the major role in the behaviour of the integrated events is played by the values of $M_{\eta^\pm}$ and the mass splitting between $\eta^\pm$ and $S$. Starting from the top-left frame, we see that a scenario with a relatively light $\eta^\pm$ produces a considerable amount of events only if the splitting is large ($\sim 350$ events with $M_{\eta^\pm}=100$ and a splitting of $30$ GeV), whereas it rapidly scales down if the splitting is set to $10$ GeV or below.

Then, we must understand what happens when the mass of $\eta^\pm$ and the splitting are increased (i.e., $M_S$ is kept fixed). We have already seen that the total cross section drops as the masses increase (see Fig.~\ref{Fig:sigma-eta-S}), but we have to remark that strong cuts on both the transverse momentum and the missing transverse energy are applied. From a kinematic point of view, we note that as more energy is released in the decay of the $\eta^\pm$, the jet(s) coming from the $W$ will be more energetic, and lead to events with more high transverse $p^T_j$ events being produced, giving rise to the behaviour that we observe in the top-right frame.

Thereafter, the same argument applies to the bottom-left frame, however we have to remark that the total cross-section is dropping with the higher $\eta^\pm$ mass and the two effects balance the production of high transverse $p^T_j$ jets.

In the bottom-right frame we can appreciate a different effect: the splitting becomes so large that the $W^\pm$ is produced on-shell. Despite the lower total cross section, the on-shell-ness of the charged gauge boson has a major role in the production of high transverse $p^T_j$ jets, so that the trend is reversed: as soon as the on-shell-ness is realised, the bigger is the splitting and the lower is the number of integrated events. For the highest-splitting case ($M_{\eta^\pm}=160~\text{GeV}$ and $M_S=70~\text{GeV}$), since the $W$ is on-shell, it decays to two ``soft'' jets, rather than a more collimated pair of merged jets.

We summarise the values of the number of integrated events for the aforementioned cases in Table~\ref{tab:event_jet}.

\begin{table}[ht]
\begin{center}
\begin{tabular}{|c|c|c|c|}
\hline
\ & $M_S=70$ GeV & $M_S=80$ GeV  & $M_S=90$ GeV \\
\hline
 $M_{\eta^\pm}=100$ GeV & $355.5$ & $73.4$ & $1.4$ \\
 $M_{\eta^\pm}=120$ GeV & $880.6$ & $523.8$ & $204.5$ \\
 $M_{\eta^\pm}=140$ GeV & $776.2$ & $697.1$ & $472.1$ \\
 $M_{\eta^\pm}=160$ GeV & $591.4$ & $660.0$ & $671.8$ \\
\hline
\end{tabular}
\end{center}
\caption{Number of integrated events at $\sqrt{s}=14$ TeV and $\Lumint=100$ fb$^{-1}$ for the process $pp\to j+\text{MET}$. Standard kinematic cuts are applied, plus a further cut requiring that $\text{MET}>120$ GeV and $p^T_j>120$ GeV.
\label{tab:event_jet}}
\end{table}

Finally, let us comment on the typical background for such process. When $W^\pm$ is off-shell, the main source of SM background is produced by the process $pp\to j+Z$ (with $Z\to 2\nu$). It is known (see \cite{ATLAS-CONF-2012-147} and references therein) that the $j+Z$ final state is produced with a very high cross section at the LHC, and the number of events is overwhelming with respect to the signal that we have profiled in this section. Indeed, we have shown that a signal originating from the charged scalar of an inert doublet can produce a sizable number of events and we have also characterised the profile of such a signal. Since the background can be well above the signal (with $10^3-10^4$ events at the peak against our $10-10^2$), a possible detection could only occur together with a dedicated analysis of the emerging jet structure \cite{ATLAS:2012am}. When $W^\pm$ is on-shell, the scenario changes significantly: requiring $W^\pm$ reconstruction clears all the background produced by the previous source. In this case, the source of SM background is the process $pp\to W+Z$ (with $Z\to 2\nu$) which produces a considerably smaller number of events ($N_\text{BG}\simeq 3000$ at $14$ TeV and $\Lumint=100$ fb$^{-1}$, with the aforementioned cuts applied). From Table~\ref{tab:event_jet}, we extract the number of events that the signal produces in such a scenario, which is $N_{S}\simeq 600$. A simple calculation leads to the conclusion that the Gaussian significance (defined as $N_S/\sqrt{N_{BG}}$) is $~10$, i.e. well above the discovery threshold. It is straightforward to estimate the required luminosity to probe a signal with a $\Sigma=5$ significance:
\begin{eqnarray}
\Lumint^\text{crit}\simeq \left(\frac{5\sqrt{N_{BG}}}{N_S}\right)^2\Lumint\simeq 20~\text{fb}^{-1},
\end{eqnarray}
which corresponds to the early stage of data collection in the high-energy scenario.

Let us also comment on another possible source of background, which is represented by the analogous process:
\begin{eqnarray}
pp\to AS\to ZSS,
\end{eqnarray}
which also leads to an effective $j+\text{MET}$ final state. Unless the $Z$ decays leptonically, these events will add to the previous ones, and can not be distinguished from the signal. It is not a negligible contribution. However, one could imagine isolating this contribution at a future $e^+e^-$ collider:
\begin{eqnarray}
e^+e^-\to AS\to ZSS.
\end{eqnarray}
It should be accessible at both the ILC \cite{Djouadi:2007ik} and CLIC \cite{Linssen:2012hp}. Hence, we conclude the section by keeping the focus on the single-$\eta^\pm$ production and don't investigate the single-$A$ production any further.

\subsection{Tracks and kinks}

\begin{figure}[htb]
 \begin{center}
\includegraphics[scale=0.43]{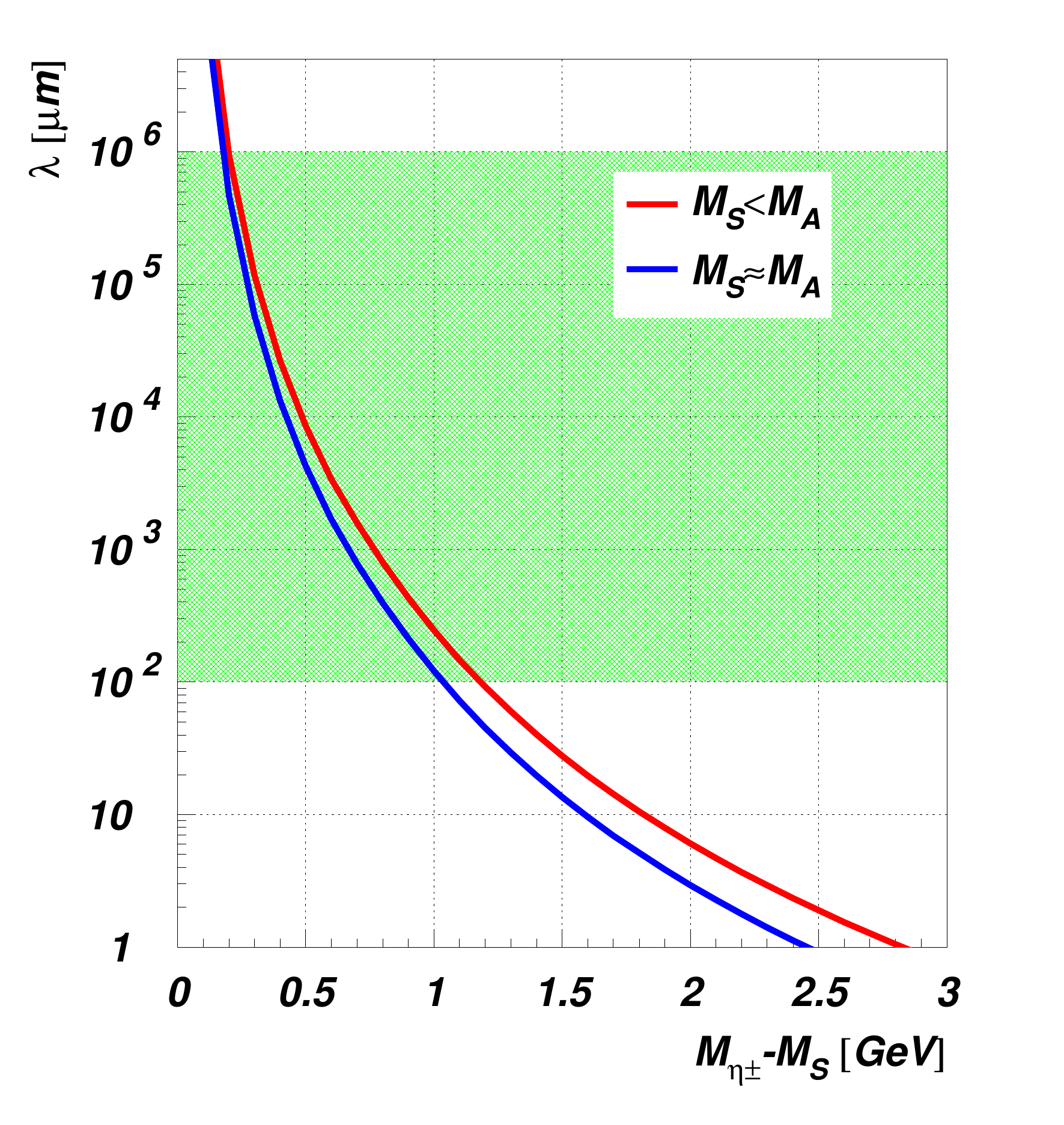}
\end{center}
\caption{Decay length $\lambda$ vs mass splitting $M_{\eta^\pm}$--$M_S$ for two relative $S$--$A$ spectra.}
\label{Fig:decay-length}
\end{figure}

As mentioned above, if the mass splitting is small, then the $\eta^\pm$ could live long enough to leave a track in the detector, before decaying. After decay, which will involve a virtual $W$, there will either be two jets, or a charged lepton and a neutrino. Thus, this secondary vertex will be the origin of two jets (merged in an effective single jet, as we have seen in the previous subsection) or a charged lepton. In each case, a displaced track (or tracks) will be present. If this $\eta^\pm$-track is longer than around $100~\mu\text{m}$, but not more than 1~m, such a track and the kink could be seen in the detector as a displaced interaction vertex. We remark that this could be the only way to detect the single-lepton signal.

The lifetime is determined by the mass splitting $M_{\eta^\pm}-M_S$. The corresponding decay length, $\lambda=c\tau$, is given in Fig.~\ref{Fig:decay-length}. The shaded region corresponds to $100~\mu\text{m}<\lambda<1~\text{m}$, which is the typical range of a micro-vertex detector. It is seen that the masses should differ by not more than 1~GeV for this to be relevant.

\section{Summary}
\label{Sec:summary}
\setcounter{equation}{0}

We have reviewed a CP-violating extension of the Inert Doublet Model. CP-violation is introduced by adding a second Higgs doublet to the IDM. The model thus has a pair of charged scalars in addition to the charged Higgs bosons. In contrast to the charged Higgs bosons, these additional charged scalars do not couple directly to fermions, but can be produced at the LHC via their couplings to the gauge bosons and neutral Higgs particles. Because of the imposed $Z_2$ symmetry (which makes the lightest member of the ``inert'' doublet a dark matter candidate), the charged scalar can only be produced in association with a neutral one, or in pairs.

Allowed parameter regions of the model have been identified. These regions are likely to be further reduceed by more precise results from the LHC. In particular, the additional charged scalar reduces the $H_1\to\gamma\gamma$ rate. Thus, if more data should reveal that $R_{\gamma\gamma}$ is signifycantly above unity, the allowed parameter space would be much reduced.

These charged scalars, if they have a mass of the order of 100~GeV, can be produced at the LHC with a sizable cross section. Their detection is however difficult, and depends on the mass splitting with respect to the dark-matter particle $S$. If the mass splitting is below a couple of GeV, then the lifetime will be long enough to yield displaced vertices in the detector. If the mass splitting exceeds the $W$ mass, then one may be able to identify the hadronic decay of the $W$ from $\eta^\pm\to SW$. For intermediate mass splittings, the off-shell $W$ will decay to two jets that will be seen as a broad unresolved jet.
Since the hadronic environment of the LHC introduces an enormous amount of background events containing a single jet plus MET, in this parameter region it is not possible to resolve the signal in a pure counting experiment and a successful phenomenological analysis must exploit more refined techniques like sophisticated jet substructure studies, combined $p_T(\text{jet})$ and MET analysis, etc. In this difficult scenario, it is important to understand if an extra jet giving rise to an effective di-jet plus MET signal could improve the quality of the analysis.

Of course, one should realize that the LHC could be insufficient to perform a satisfactory search for $\eta^\pm$ particles. In principle, a future linear collider like the ILC or CLIC could help to complete the information on the parameter space by allowing a study of the twin channel $e^+e^-\to AS\to ZSS\to j+\text{MET}$.


\subsubsection*{Acknowledgements}  
\noindent 
The work of PO and MP has been supported by the Research Council of Norway.
The work of GMP has been supported by the German Research Foundation DFG through Grant No.\ STO876/2-1 and by BMBF Grant No.\ 05H09ODE. 
The work of AP has been supported by the grant  RFBR 12-02-93108-CNRSLa.
GMP and AP would like to thank the Galileo Galilei Institute (GGI) in Florence for the kind hospitality while part of this work was carried out. GMP is also grateful to the INFN (Sezione di Cagliari) for the hospitality and logistical support during the completion of this work.


\end {document}